\documentclass[rnote]{aa}
\usepackage[pdftex]{graphicx}
\usepackage{txfonts}
\usepackage{natbib}

\begin{document}

   \title{Shock-related radio emission during coronal mass ejection lift-off?}

   \author{S. Pohjolainen}

   \institute{Tuorla Observatory/Department of Physics, University of Turku,
              21500 Piikki\"o, Finland\\
              \email{silpoh@utu.fi}
                   }

   \date{}

   \abstract{}{We identify the source of fast-drifting decimetric--metric 
               radio emission that is sometimes observed prior to the 
               so-called flare continuum emission. Fast-drift structures 
               and continuum bursts are also observed in association with 
               coronal mass ejections (CMEs), not only flares.}
              {We analyse radio spectral features and images acquired
                at radio, H$\alpha$, EUV, and soft X-ray wavelengths, 
                during an event close to the solar limb on 2 June 2003.}
              {The fast-drifting decimetric--metric radio burst corresponds 
                to a moving, wide emission front in the radio images, 
                which is normally interpreted as 
                a signature of a propagating shock wave. 
                A decimetric--metric type II burst where only the second 
                harmonic lane is visible could explain the observations. 
                After long-lasting activity in the active region, the hot and 
                dense loops could be absorbing or suppressing emission at the 
                fundamental plasma frequency. 
                The observed burst speed suggests a super-Alfv\'enic 
                velocity for the burst driver. The expanding and opening 
                loops, associated with the flare and the early phase of CME 
                lift-off, could be driving the shock. Alternatively, an
                instantaneous but fast loop expansion could initiate a freely 
                propagating shock wave. 
                The later, complex-looking decametre--hectometre wave 
                type III bursts indicate the existence of a propagating shock, 
                although no interplanetary type II burst was observed during 
                the event. The data does not support CME bow shock or a shock 
                at the flanks of the CME as the origin of the fast-drift 
                decimetric--metric radio source. Therefore 
                super-Alfv\'enic loop expansion is the best candidate for the 
                initiation of the shock wave, and this result challenges the 
                current view of metric/coronal shocks originating either in  
                the flanks of CMEs or from flare blast waves. 
                }
                {}

   \maketitle


\section{Introduction}

Frequency-drifting features at decimetric--metric wavelengths are usually
observed in association with flares and coronal mass ejections (CMEs).
Some groups of fast-drift bursts within a restricted bandwidth have been 
identified as type II precursors \citep{klassen99}, and interpreted as 
signatures of reconnection processes above expanding soft X-ray loops that 
later lead to type II burst emission \citep{klassen03}. 
Type II bursts are generally believed to be formed by propagating shock 
fronts that accelerate electrons; these electrons excite Langmuir waves, 
which convert to radio waves observable as plasma emission close to the 
fundamental and second-harmonic frequencies \citep{melrose80,cairns03}. 
As the burst exciter moves outwards in the solar atmosphere, the plasma 
density decreases, which causes a frequency-drift to lower frequencies. 
The highest frequencies of the fundamental components of type II bursts 
rarely exceed a few hundred MHz \citep{lin06}. 

Shocks can be either freely propagating blast waves or driven shocks 
(piston-driven or bow-shocks ahead of a projectile); for a recent review, 
see \citet{warmuth07}. There is an ongoing debate about whether the 
exciters of metric type II bursts are flare-related or CME-related 
(e.g., Cane \& Erickson, \citeyear{cane05}). Metric type II bursts are 
sometimes accompanied by stationary and long-lasting continuum emission, 
which forms out of the trailing edge of a type II burst \citep{robinson75}. 
These have been classified as type II-related flare continuum bursts (FCII), 
and the emission is most probably plasma radiation at the fundamental 
frequency. 
Flare continuum (FC) emission was introduced by \citet{wild70}, 
to separate the strong, early-in-the-flare appearing continuum from the 
stationary type IV continuum that usually appears well after the flare 
or after the lift-off phase of a CME. To add confusion to this classification 
scheme, flare continuum can sometimes be followed by a moving type IV burst 
(i.e., where the source is moving spatially, not just frequency-drifting);
a schematic representation is provided by \citet{robinson78}. 

\citet{vourlidas07} presented an event that shows a long-duration continuum 
burst at 70--20 MHz. This continuum feature fitted well into the FCII 
classification, since it was associated with a type II burst. 
\citet{farnik03} had already analysed the earlier evolution of the
same event, and observed an ``unusual drifting continuum'' in the 
400\,--\,40 MHz frequency range, preceding the FCII continuum burst  
(see Fig. 14 of their paper). They associate the unusual drifting metric 
continuum with expanding EUV and soft X-ray loops.

\begin{figure*}[!ht]
    \includegraphics[width=12cm]{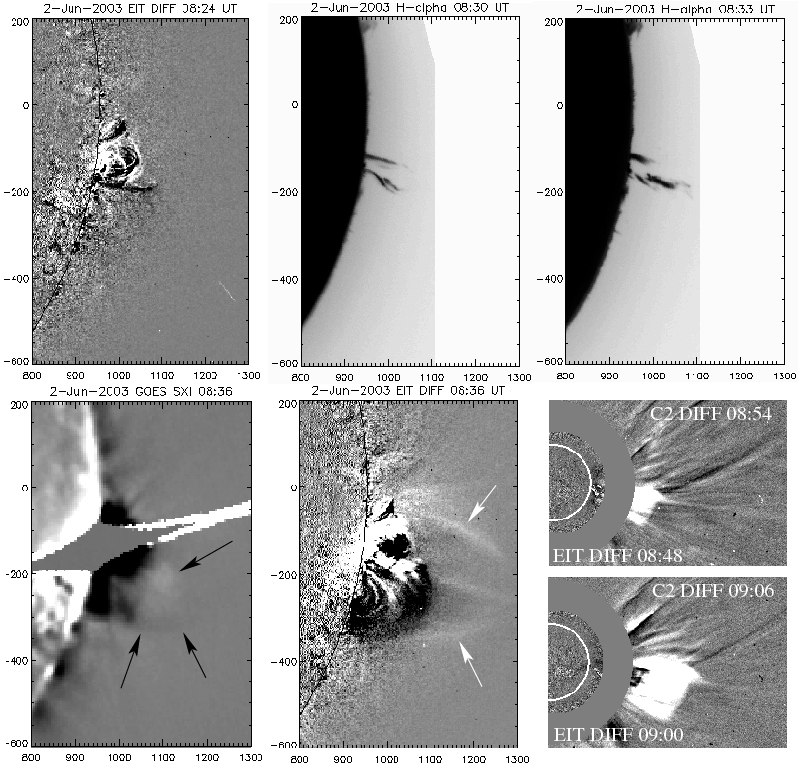}
    \caption{Evolution of the 2 June 2003 eruption: SOHO EIT difference 
     image at 08:24--08:12 UT, Kanzelh\"ohe H$\alpha$ images at 08:30  
     and 08:33 UT, GOES SXI difference image at 08:36--08:32 UT,
     SOHO EIT difference image at 08:36--08:12 UT, and SOHO LASCO 
     C2 difference images at 08:54 and 09:06 UT (LASCO CME 
     Catalog). At 08:54 UT, the CME front is already at heliocentric 
     distance 3.1 $R_{\odot}$.
     At 08:36 UT, a bright loop-like structure is observed in soft X-rays 
     (indicated with black arrows) and at the same time bright EUV rays  
     are visible over the active region (white arrows). 
     The hard X-ray experiment RHESSI observed an emission source just 
     above the solar limb at 08:14 UT, but unfortunately the satellite 
     entered night-time one minute later, and no comparison could be
     made with the hard X-ray data. An H$\alpha$ flare was recorded at 
     08:41--08:44 UT (NGDC NOAA flare lists, data not shown here).    
     }
     \label{fig1}
\end{figure*}

We analyse a rare event with a continuum similar to that of a FCII 
classification, apart from the fact that no interplanetary type II burst 
is detected. The FCII-like structure was preceded by an unusual 
fast-drifting decimetric--metric radio feature that started at a 
very high frequency, close to 1 GHz, and later blended with the FCII-like 
continuum. Since this event occurred close to the solar limb, and imaging 
at radio, EUV, H$\alpha$, and soft X-ray wavelengths exist, we attempt 
to determine the exciter of the fast-drift feature, and to ascertain 
whether this feature and the FCII-like continuum are related. We 
also consider the connection with the simultaneously appearing CME.

\section{Observations}

A GOES M3.9 class flare in NOAA AR 10365 (S07\,W89) on 2 June 2003 started 
with a flux rise at 08:12 UT, followed by new rises at 08:21 and 08:27 UT, 
before the flux maximum at 08:37 UT. Since this was a limb flare, it is 
possible that the flare was partly occulted and hence the start time and 
flux evolution are uncertain. A filament eruption started in between the 
available Kanzelh\"ohe H$\alpha$ images at 08:23 and 08:30 UT. From later 
images, it is possible to estimate the (projected) plane-of-the-sky speed 
of the filament, which is at least 330 km s$^{-1}$. 

SOHO EIT \citep{boudin95} images, starting at 08:24 UT, show the erupting 
filament, together with loop displacements. In the EIT images at 08:36 and 
08:48 UT large-scale changes in the active region are evident. At 08:36 UT, 
bright EUV structures are observed at height $\approx$1.35 $R_{\odot}$, 
which is higher than the simultaneously observed soft X-ray loop-like 
structure, observed by GOES SXI near 1.27 $R_{\odot}$, see Fig. \ref{fig1}.  

A partial halo-type CME was first observed at 08:54 UT by SOHO LASCO
\citep{brueckner95}, close to a heliocentric height of 3.1 $R_{\odot}$ 
(Fig. \ref{fig1}). The CME velocity, measured using a linear fit to all 
height--time data points, was 980 km s$^{-1}$ (LASCO CME Catalog). 
No strong evidence of acceleration or deceleration in the CME speed 
was observed. Backward-extrapolation of the CME heights places the CME 
front close to the solar limb at about 08:30 UT.

\begin{figure*}
  \centering
    \includegraphics[width=16cm]{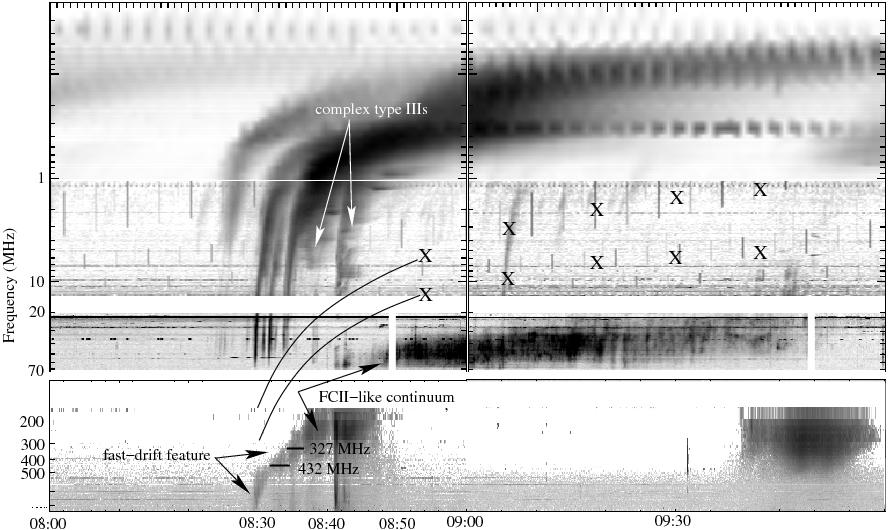}
    \caption{A fast-drift radio feature was observed on 2 June 2003 from 
             1000 MHz to around 300 MHz, where the emission lane 
             blended with a wide-band FCII-like continuum emission. The 
             continuum feature was best observed at 70--20 MHz, and it 
             decreased in intensity close to 10 MHz at about 09:50 UT. 
             Complex-looking type III bursts were observed after 08:36 UT.   
             The ``X''s mark the plasma frequencies calculated from the 
             observed heights of the plane-of-the-sky CME front locations, 
             using Saito and 2-times Newkirk atmospheric density models. 
             Solid black lines represent the frequency--time profiles for a 
             disturbance travelling at a speed of 980 km s$^{-1}$. This 
             spectral composite includes observations from Phoenix-2,
             Nan\c{c}ay Decameter Array, and Wind WAVES. 
}
     \label{fig2}
\end{figure*}

The radio dynamic spectrum from Phoenix-2 \citep{messmer99}
shows the onset of a fast-drift burst close to 1000 MHz at 08:29 UT 
(Fig. \ref{fig2}). 
The frequency drift rate was initially about 2 MHz s$^{-1}$, but then 
decreased to a constant value of about 0.7 MHz s$^{-1}$, between 08:32
UT and 08:37 UT.
The narrow emission band and the fast drift suggest that this is plasma 
radiation. The local plasma frequency (fundamental emission), $f_p$, can 
be used to estimate the local electron density, $n_e$, as 
$f_p$\,=\,9000$\sqrt{n_e}$. Electron density can in turn be converted to 
height with the use of atmospheric density models. The use of density 
models, and the uncertainties involved, are explained in detail in, 
e.g., \citet{pohjolainen07}. 

\begin{figure*}
    \includegraphics[width=12cm]{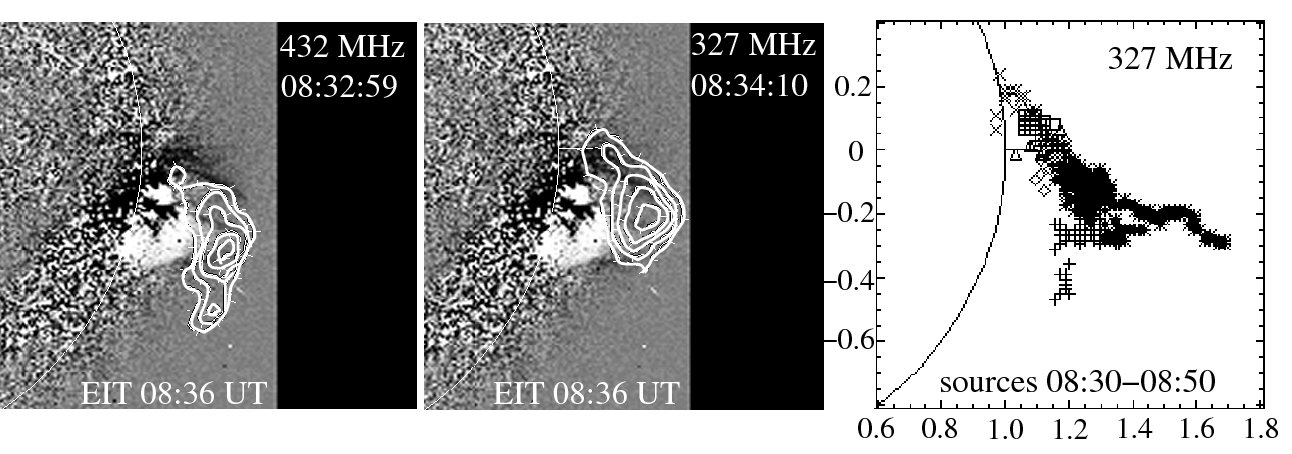}
    \caption{Contours of the radio emission sources at 08:32:59 UT 
             (432 MHz, left) and 08:34:10 UT (327 MHz, middle), observed 
             during the fast-drift spectral feature (Nancay Radioheliograph 
             imaging). 
             The radio source centroids during 08:30\,--\,08:50 UT at 327 MHz 
             (right) show both the fast-drift feature and the FCII-like 
             continuum sources. The EUV image on the background is the EIT 
             difference image with reversed color scale at 08:36 UT
             (the bright EUV rays show as black).
             The EIT field of view does not extend beyond 1.4 R$_{\odot}$.
             The three plots all show the same region, and the axes are  
             in solar radiae.}
     \label{fig3}
\end{figure*}

At 08:31:10 UT, the low-frequency edge of the 
fast-drift emission lane was located at 500 MHz, which corresponds to an 
electron density of 3.1$\times$10$^9$ cm$^{-3}$ (assuming fundamental 
emission), and by 08:35:00 UT the emission had drifted to 300 MHz 
($n_e$ $\approx$ 1.1$\times$10$^9$ cm$^{-3}$). 
We assume that the density gradient followed standard atmospheric models, 
because the frequency drift was almost constant during this time.
Since the highest densities are found only in active region loops and 
streamers, we have 
to multiply the standard atmospheric models by suitable coefficients. 
To determine these coefficients, we use observed radio source heights 
as a constraint.  The fast-drift emission sources were imaged at the 
Nan\c{c}ay Radioheliograph  \citep{alain} frequencies of 432 and 327 MHz. 
The radio images show a wide arc-like front over the western limb, 
above 1.2 $R_{\odot}$ after 08:32 UT (Fig. \ref{fig3}).
To reproduce the observed height of 1.2 $R_{\odot}$ at 327 MHz, we 
need to use 19-times Saito \citep{saito70} or 8-times Newkirk 
\citep{newkirk61} model densities. A 19-times Saito model density gives 
a height of 1.097 R$_{\odot}$ for 500 MHz, and a height of 
1.231 R$_{\odot}$ for 300 MHz. If we calculate the speed directly from 
these heights and times, it is about 405 km s$^{-1}$.

The direction of the moving radio arc was towards the Northwest, 
and it appears as if the front was moving away from the observer.  
The observed speed must therefore be corrected for projection effects. 
The observed, projected velocity of the front was approximately 300 
km s$^{-1}$, suggesting a ``true'' deprojected velocity of 
400\,--\,500 km s$^{-1}$. 

At about 08:36 UT, the emission lane of the fast-drift feature disappeared 
into a wider continuum emission, starting from close to 300 MHz. 
A slowly-moving emission source is observed in the Nancay Radioheliograph 
images at 327\,--\,150 MHz, during 08:42--09:00 UT (Fig. \ref{fig3}). 
These frequencies are close to the high-frequency edge of the continuum, 
which was observed best at 70--20 MHz (Nan\c{c}ay Decameter Array 
observations; Lecacheux, \citeyear{lecacheux00}). The continuum emission 
faded away from the spectrum at about 09:50 UT, close to 10 MHz (Wind 
WAVES observations; Bougeret et al., \citeyear{bougeret95}).
 
Figure \ref{fig2} shows the estimated emission frequencies at the CME 
front, for the case in which the CME creates a bow shock at its leading 
edge. We note, however, that no interplanetary type II burst was visible at 
decametre-hectometre wavelengths. The frequencies were calculated using 
the standard atmospheric density models of Saito and 2-times Newkirk; 
the use of even higher density models would move the "X"s and the 
frequency-drift profiles to lower positions in the plot, towards higher 
frequencies. 

Wind WAVES observations show complex-looking radio type III
bursts starting around 08:36 UT (Fig. \ref{fig2}). The tilted and
disturbed type III burst lanes have earlier been associated with 
shocks, either by electron beams traversing turbulent shock regions 
\citep{reiner00, lehtinen08} or by electrons accelerated by the shock  
\citep{cane81}. Either way, and in spite of the fact that no interplanetary
type II burst was observed, a propagating shock must have existed.

\section{Results and Discussion}

In the analysed event, as in the event described by \citet{farnik03},
the start of radio emission was preceded by long-term (tens of minutes) 
activity in X-rays. The outcome of earlier heating would have been that 
hot and dense loops existed high in the corona. The observed plasma 
emission would reflect the high densities in these loops, and the
high densities explain the high values of starting frequency. The 
initially fast drift rates could be attributed to the rapidly-changing 
densities. Radio images of the fast-drift spectral feature 
show an outward-moving arc-like structure. This structure is observed 
above the limb at heights $>$1.2 R$_{\odot}$, which is much higher than 
predicted by any standard atmospheric density model, at these frequencies.

The frequency drifts of metric type II bursts are usually in the range 
of 0.1\,--\,1.0 MHz s$^{-1}$ \citep{nelson85}. The observed frequency drift
rate of the fast-drift feature was $\approx$0.7 MHz s$^{-1}$ between 500 
and 300 MHz, after the very fast start. The duration of this spectral  
feature was 6\,--\,7 minutes, which is well within the observed usual 
durations of metric type II bursts. It is also known that due to plasma 
processes, emission at the fundamental plasma frequency can be weaker than 
at the second harmonic at metric wavelengths. For decametre--hectometre 
waves, the converse occurs. Thus, it is possible that the fast-drift 
metric radio feature was the second harmonic lane of a type II burst, and 
due to some (unknown) processes the fundamental emission band was either 
suppressed or absorbed. Of course, the non-visibility of weaker fundamental 
lanes can be due to instrumental effects.
By adopting the type II interpretation, the fast-drift feature 
fulfils the definition of a later-appearing FCII continuum, because then
it is type II burst-related. The tilted and complex-looking type III bursts 
also indicate the presence of a shock wave.
 
If we assume that the observed emission was emission at the second harmonic, 
the corresponding fundamental plasma frequencies are 250 MHz and 150 MHz. 
Using generally-accepted 10-times Saito model densities, these 
frequencies provide atmospheric heights of 1.19 and 1.37 R$_{\odot}$, 
respectively. These values are close to the observed source heights and 
imply a burst speed of 545 km s$^{-1}$. 
The calculations in Section 2 provide a burst speed of 405 km s$^{-1}$;
to obtain this value, we need to assume 19-times Saito atmospheric 
densities if emission occurs at the fundamental plasma frequency.
Both of these values are still in agreement with the 400\,--\,500 
km s$^{-1}$ de-projected speed of the imaged radio arc.

What is the origin of the shock responsible for the type II emission?
In the vicinity and high above active regions, where the densities are high 
but the magnetic field strengths are low, the local magnetosonic (Alfv\'en) 
speed can be as low as 200 km s$^{-1}$; see for example \citet{warmuth05}. 
A super-Alfv\'enic shock is then possible. Ejected material or 
high-speed loop expansion could drive the shock. Freely-propagating blast 
waves could accelerate the electrons required to produce a type II burst, 
although these waves would die out sooner than the driven, 
super-Alfv\'enic waves. 
The bright EUV rays, above the soft X-ray loop, alludes to an overlying 
arcade that opens during CME formation; the rays would then represent 
compressed material at the flanks. However, a shock formed at the flank 
of a CME would be unlikely to display a wide, moving radio arc.
No H$\alpha$ Moreton wave (a blast wave signature) was visible 
during this event, although at the limb they are difficult to detect. 
No loop ejection was observed either. Therefore super-Alfv\'enic loop 
expansion is the best candidate for the initiation of the shock wave, 
and this result challenges the current view that metric/coronal shocks 
originate either at the flanks of CMEs or from flare blast waves. 

The frequency--time profiles for the later-observed white-light CME 
front do not agree with the evolution of the fast-drift decimetric--metric
feature. The frequency--time track of the CME is, however, similar to 
the frequency drift of the FCII-like continuum. \citet{kai69} reported 
an event in which the radio images at 80 MHz show arc-like radio emission, 
followed closely by an outward-moving single source. The single moving 
source was then classified as a moving type IV burst. Since our FCII-like 
continuum source shows movement outwards, the assumption that it is  
a "stationary flare continuum source" can also be questioned.  

Since the analysis of high-frequency, fast-drift features requires 
both wide frequency range spectral and radio imaging observations 
combined with multi-wavelength high time-cadence imaging, the results 
presented here are preliminary and require comparison to other similar
-- but rare -- events.

\begin{acknowledgements}
SOHO is a project of international cooperation between ESA and NASA.
GOES SXI is operated by the National Oceanic and Atmospheric Administration
and their data is available at the NGDC website. The LASCO CME Catalog 
is generated and maintained at the CDAW Data Center by NASA and The 
Catholic University of America in cooperation with the Naval Research 
Laboratory, and it is available at 
\texttt{http://cdaw.gsfc.nasa.gov/CME\_list/}. 
Radio data was accessed with the help of the Radio Monitoring 
Survey at \texttt{http://secchirh.obspm.fr/select.php}, 
generated and maintained at the Observatoire de Paris by the 
LESIA UMR CNRS 8109 in cooperation with the Artemis team, Universities 
of Athens and Ioanina and the Naval Research Laboratory. 
The solar radio group at LESIA is thanked for the Nancay Radioheliograph 
data. The Phoenix-2 dynamic radio spectral data are available at the 
ETHZ Radio Astronomy and Plasma Physics group website.  Thanks are due to 
Vasyl Yurchyshyn for providing the H$\alpha$ data (Global High-Resolution 
H$\alpha$ Network, of which Kanzelh\"ohe Observatory is a member). The 
comments and suggestions from the referee improved the paper significantly. 
Rami Vainio is thanked for discussions and careful reading of the 
manuscript.  
\end{acknowledgements}

{}

\end{document}